\documentclass[baaa]{baaa}

\usepackage[pdftex]{hyperref}
\usepackage{subfigure}
\usepackage{natbib}
\usepackage{helvet,soul}
\usepackage[font=small]{caption}
\usepackage{float}
\usepackage{nccmath}

\contriblanguage{1}

\contribtype{2}

\thematicarea{3}

\title{Simulations of fully convective M dwarfs: dynamo action with varying magnetic Prandtl numbers}

\titlerunning{Simulations of fully convective M dwarfs}

\author{
C.A. Ortiz-Rodr\'iguez\inst{1}, 
D.R.G Schleicher\inst{1},
P.J. K\"apyl\"a\inst{2,3} \&
F.H. Navarrete\inst{4,3}
}
\authorrunning{Ortiz-Rodr\'iguez et al.}

\contact{ortizrodcarolina@gmail.com}

\institute{
Departamento de Astronom{\'\i}a, Universidad de Concepci\'on, Chile
\and
Institut f\"ur Astrophysik, Georg-August-Universit\"at G\"ottingen, Alemania
\and
Nordita, KTH Royal Institute of Technology and Stockholm University, Suecia
\and
Hamburger Sternwarte, Universit\"at Hamburg, Alemania
}

\resumen{Las enanas M son estrellas de secuencia principal de baja masa, el tipo de estrella m\'as numeroso en el vecindario solar conocidas por tener actividad magn\'etica. El objetivo de este trabajo es explorar las soluciones de dinamo y los campos magn\'eticos de enanas M completamente convectivas con diferentes numeros magnético de Prandtl ${\rm Pr_M}$, y con un período de rotaci\'on ($\rm P_{rot}$) intermedio de 43 d\'ias. Se sabe que ${\rm Pr_M}$ juega un papel importante en la acción del d\'inamo; d\'inamos para bajos $\rm Pr_M$ y altos $\rm Pr_M$ tienen propiedades muy diferentes. Usamos simulaciones num\'ericas magnetohidrodin\'amicas (MHD) tridimensionales con el modelo ``star-in-a-box'' usando par\'ametros estelares para una enana M5 con $0.21 M_{\odot}$. Encontramos que las soluciones del d\'inamo son sensitivas al $\rm Pr_M$, que es equivalente al n\'umero magn\'etico de Reynolds ${\rm Re_M}$. Las simulaciones con este periodo de rotación presentan ciclos per\'iodicos del campo magn\'etico a gran escala para ${\rm Pr_M} \leq 2$; para valores m\'as altos los ciclos desaparecen y a muestran inversiones irregulares del campo. Nuestros resultados son consistentes con estudios previos y sugiere que los d\'inamos que operan en estrellas completamente convectivas se comportan de manera similar a los que operan en estrellas parcialmente convectivas.}

\abstract{M dwarfs are low-mass main-sequence stars, the most numerous type of stars in the solar neighbourhood, which are known to have significant magnetic activity. The aim of this work is to explore the dynamo solutions and magnetic fields of fully convective M dwarfs with varying magnetic Prandtl numbers ${\rm Pr_M}$, and a rotation period (${\rm P_{rot}}$) of 43 days. ${\rm Pr_M}$ is known to play an important role in the dynamo action; dynamos for low-$\rm Pr_M$ and large-$\rm Pr_M$ have very different properties. We performed three-dimensional magnetohydrodynamical (MHD) numerical simulations with the ``star-in-a-box'' model using stellar parameters for an M5 dwarf with $0.21 M_{\odot}$. We found that the dynamo solutions are sensitive to ${\rm Pr_M}$. The simulations at this rotation period present periodic cycles of the large-scale magnetic field up to ${\rm Pr_M} \leq 2$; for higher values the cycles disappear and irregular solutions to arise.  Our results are consistent with previous studies and suggest that the dynamos operating in fully convective stars behave similarly as those in partially convective stars.}


\begin{document}

\maketitle
\section{Introduction}\label{S_intro}
Many type of stars across the Hertzsprung-Russel diagram are known to have magnetic fields. M dwarf stars are low-mass late-type main-sequence stars that are known for being the most abundant type of stars in our galaxy and for exhibiting significant surface magnetic activity, as has been reported by several authors (see \cite{kochukhov2021magnetic} and references therein). Their stellar structure is determined by the mass, which ranges from $0.08$ to $0.55\, M_{\odot}$;  M dwarfs with masses greater than $0.35\, M_{\odot}$ are partially convective (solar-like),  which means they have a radiative core and a convective envelope, whereas M dwarfs with masses less than $0.35 M_{\odot}$ are fully convective \citep{Chabrier&Baraffe1997}. 

Stellar magnetic fields are sustained by a dynamo mechanism that works in the convective zone that, basically, converts the kinetic energy into magnetic energy \citep{Brandenburg_2005}. Fully convective M dwarfs do not posses a tachocline, which is the shear layer between the radiative and convective zone in solar-like stars and the importance of which for stellar dynamos is under debate. Since it is not possible to observe directly the magnetic field in the interior of stars, three-dimensional numerical simulations of stars are performed with the aim of achieving a better understanding of their magnetic fields, dynamos, and convection as functions of stellar parameters, like mass, age, rotation and dimensionless numbers that describe the physics in the stellar interiors. As a consequence, in this work we present three-dimensional magnetohydrodynamical numerical simulations of fully convective M dwarf stars at intermediate rotation period, varying the magnetic Prandtl number between 0.1 and 10. The magnetic Prandtl number, $\rm Pr_M$, is the ratio of kinematic viscosity ($\nu$) to magnetic diffusivity ($\eta$) of the fluid. Thereby, the magnetic Prandtl number is an intrinsic property of the fluid and has an important effect on the dynamo. Because this dimensionless parameter is proportional to temperature and inversely proportional to density, we can find a wide variety of $\rm Pr_M$ values across the universe. Galaxy clusters, for example, have temperatures in the order of $10^8$ K and very low densities ($\sim 10^{-26}$ g cm$^{-3}$), resulting in very high magnetic Prandtl numbers ($10^{29}$), whereas temperatures in the upper part of the solar convection zone are on the order of $10^{4}$ K and densities of $10^6$ g cm$^{-3}$, resulting in $Pr_{\rm M} \ll 1$. \citep{Brandenburg_2005}. This manuscript is presented as follows: section \ref{sec:methods} presents the computational methods used with a brief description of the model and the stellar parameters for an M5 star. In section \ref{sec:results} we present the main results; conclusions are given in section \ref{sec:conclusions}.

\section{Computational methods}\label{sec:methods}
\subsection{The Pencil Code}
The simulations were run with the {\sc Pencil Code}\footnote{https://github.com/pencil-code/} \citep{Collaboration2021}, which is a high-order finite-difference code for compressible hydrodynamic flows with magnetic fields that solves ordinary and partial differential equations. The code runs efficiently under MPI on massively parallel computers with distributed memory. The Pencil Code is highly modular, with physical and technical modules that can be turned on and off based on the needs of the user. 

\subsection{Star-in-a-box model}
The model used in this work is the star-in-a-box setup described in \cite{kapyla2021star}. It allows dynamo simulations of fully convective stars in a sphere of radius $R$ inside a cubic box with a side of $2.2\,R$, using a Cartesian grid. The model is described by the equations of magnetohydrodynamics (MHD).

\subsubsection{Stellar parameters}
We consider an M5 dwarf with the same stellar parameters used by \citet{kapyla2021star}, i.e. stellar mass $M_{\star} = 0.21 M_{\odot}$, radius $R_{\star} = 0.27 R_{\odot}$, and luminosity $L_{\star} = 0.008 L_{\odot}$ ($R_{\odot}$ and $L_{\odot}$ denote the solar radius and luminosity, respectively). We used an effective temperature $T_{\rm eff} =4000$ K and a central density $\rho_c^{\star} \approx 1.5 \cdot 10^{5}$ kg m$^{-3}$. These are typical values for an M5 dwarf and we have verified them using simulations with Modules for Experiments in Stellar Astrophysics (MESA; \citealt{paxton2010modules}), a one dimensional stellar evolution code to evolve single stars. Hence, for a real M5 star we have $\mathcal{L}_{\star} = 2.4 \cdot 10^{-14}$. We use here  the enhanced luminosity approach introduced by \cite{2020GApFD.114....8K}, where the gap between the shortest (acoustic timestep) and longest (Kelvin-Helmholtz time) timescales is compressed such that the latter can be resolved in the simulations. The luminosity ratio between the simulations and an M5 star is $L_{\rm ratio} = \mathcal{L}/\mathcal{L_{\star}}\approx 2.1 \cdot 10^9$. Since the enhanced luminosity leads to an increase of the convective velocity as $u_{\rm conv} \propto L_{\rm ratio}^{1/3}$, the velocities are greater by a factor of $L_{\rm ratio}^{1/3} \approx 1280$. These scaling relations have been established in previous  studies \citep{2020GApFD.114....8K, 2021arXiv210211110N}, and allow us to relate our numerical results to astrophysical stars.

\section{Results: Dynamo solutions}\label{sec:results}
\begin{figure*}[!t]
\centering
\includegraphics[width=\textwidth]{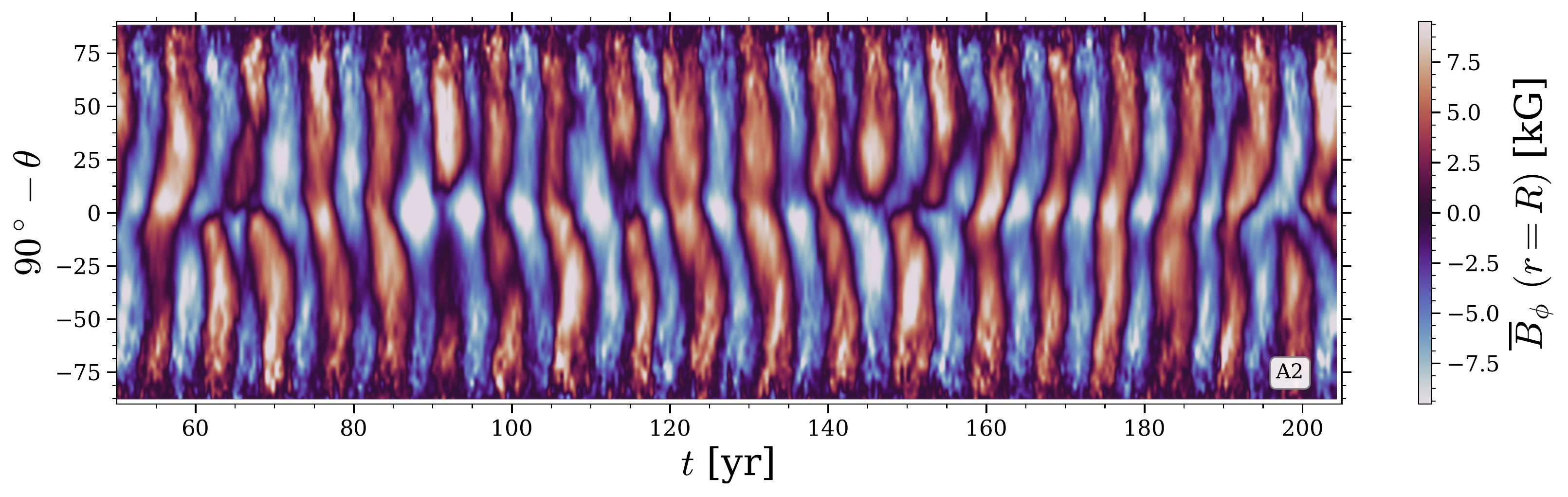} \\
\includegraphics[width=\textwidth]{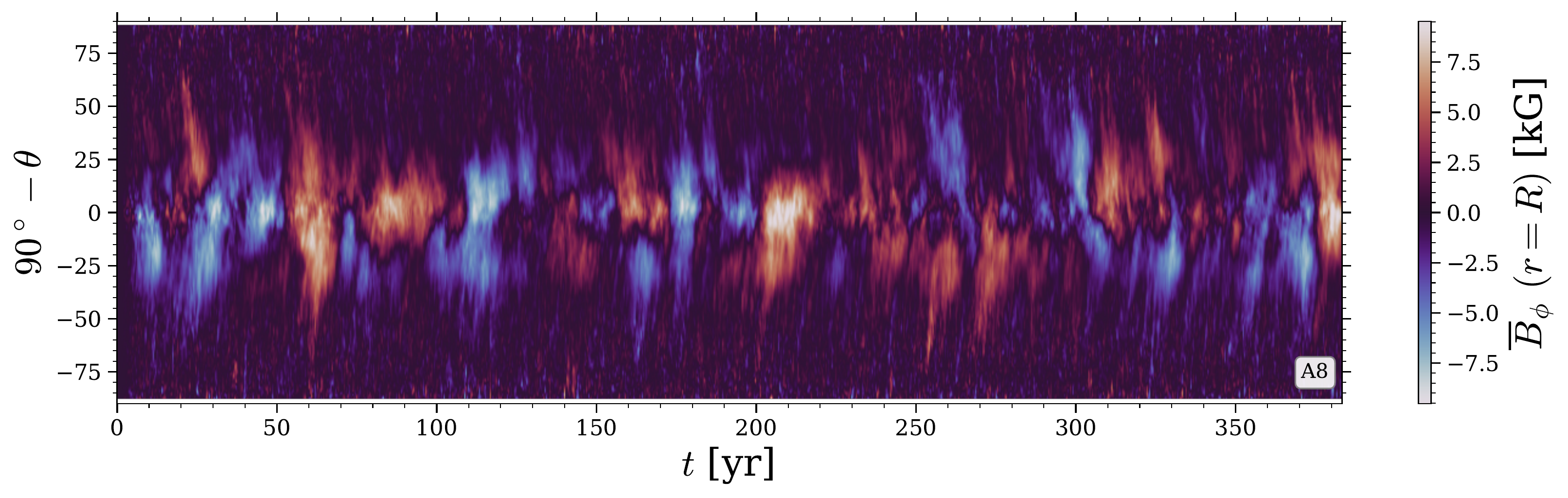}
\caption{Time evolution of the azimuthally averaged azimuthal magnetic field $\overline{B}_{\phi}(R, \theta, t)$ near the surface of the star. The colorbar indicates the strength of the magnetic field. \emph{Top (bottom) panel} is for simulation A2 (A8).} 
\label{Bpphi}
\end{figure*}

As mentioned earlier we calculated dynamo solutions at different $\rm Pr_M$ ranging from 0.1 to 10 for a rotation period $P_{\rm rot} = 43$ days. A summary of the simulations can be found in Table \ref{tabla}. For the magnetic Prandtl numbers explored in this work we find mainly two dynamo solutions: On the one hand, for $\rm Pr_M \leq 2$ the magnetic field is cyclic, similar to the butterfly diagram of the sunspots. The length of the cycles increases with increasing $\rm Pr_M$, which is equivalent to the magnetic Reynolds number. We recall that the fluid and magnetic Reynolds number are given by 
\begin{ceqn}
    \begin{align}
        {\rm Re} = \frac{u_{\rm rms}}{\nu k_{R}}, \; \; \; \; {\rm Re}_{\rm M} = \frac{u_{\rm rms}}{\eta k_R},
    \end{align}
\end{ceqn}
where $k_R = 2 \pi / R$ is scale of the largest convective eddies. The top panel in Figure \ref{Bpphi} shows the time evolution of the azimuthally averaged azimuthal magnetic field, $\overline{B}_{\phi}(R, \theta, t)$, (also known as butterfly diagram) near the surface of the star for simulation A2 with $\rm Pr_M = 0.2$, where its solution is predominantly axisymmetric and periodic cycles can be noted, which were confirmed using the Fast Fourier Transform (FFT). The length of the cycles are indicated in the last column of Table \ref{tabla}. On the other hand, the cyclic solutions start to disappear in simulations with $\rm Pr_M > 2$, presenting irregular reversals of the magnetic field. The bottom panel of Figure \ref{Bpphi} shows $\overline{B}_{\phi}(R, \theta, t)$ of simulation A8 with $\rm Pr_M = 10$, where the solution is still predominantly axisymmetric and the mean field is concentrated at mid latitudes with irregular reversals, while near the poles the mean fields are weak. All simulations presented in this work exhibit magnetic field strengths on the order of several kG. Similar solutions have been found from simulations in spherical shells by \cite{kapyla2017convection}, where the cycles increase their lengths with increasing $\rm Pr_M$ ($\rm Re_M$) until the cycles disappear in the highest $\rm Pr_M$ cases. Furthermore, in the simulations in the low-$\rm Pr_M$ regime presented (A1-A5) the dynamo waves propagate in the poleward direction at mid latitudes, while in the last three simulations (three lasts rows of Table \ref{tabla}) with the highest-$\rm Pr_M$ no clear cyclicity is detected.

\begin{table}[!t]
\centering
\caption{Summary of the simulations. From left to right the columns show the name of the run, magnetic Prandtl number, magnetic and fluid Reynolds number, volume average magnetic field strength (in kG) and volume average velocity (in m/s) and cycle periods ($\tau_{\rm cyc}$) (in years) of the large-scale magnetic field (if cyclic).  }
\begin{tabular}{ccccccc}
\hline 
\hline
Run & $\rm Pr_M$ & $\rm Re_M$ & $\rm Re$ & $B_{\rm rms}$ & $u_{\rm rms}$ & $\tau_{\rm cyc}$\\ \hline
A1  & 0.1        & 54         & 548      & 12.0                   & 10.1                & $\sim 8$                \\
A2  & 0.2        & 54         & 271      & 11.8                   & 10.1                & $\sim 8$                \\
A3  & 0.5        & 54         & 108      & 10.5                   & 10.1                & $\sim 8.5$              \\              
A4  & 1          & 73         & 73       & 10.7                   & 9.4                 & $\sim 9$                \\
A5  & 2          & 98         & 49       & 10.0                   & 9.1                 & $\sim 10.5$             \\
A6  & 5          & 208        & 41       & 11.9                   & 7.7                 & -                       \\
A7  & 7          & 299        & 41       & 11.5                   & 7.7                 & -                       \\
A8  & 10         & 388        & 39       & 11.7                   & 7.2                 & -         \\          
\hline
\end{tabular}
\label{tabla}
\end{table}

\section{Conclusions and outlook}\label{sec:conclusions}
For the rotation period of 43 days explored in this work we found mainly two dynamo solutions: First, cyclic solutions with poleward migration of activity and increasing cycle periods with higher $\rm Pr_M$ for $\rm {Pr_M} < 2$. Second, the simulations with higher $\rm Pr_M$ exhibit irregular solutions with no clear cyclicity.

It will be necessary to explore on the dynamo solutions varying $\rm Pr_M$ with different rotation rates, and to analyze the simulations in more detail. We plan in particular to explore a 90-day rotation period, which corresponds to Proxima Centauri, which have been claimed to have a magnetic field cycle of roughly seven years \citep{klein2021large}.


\begin{acknowledgement}

The authors acknowledge the Kultrun Astronomy Hybrid Cluster (projects Conicyt Programa de Astronomia Fondo Quimal QUIMAL170001, Conicyt PIA ACT172033, and Fondecyt Iniciacion 11170268) for providing HPC resources that have contributed to the research results reported in this paper. Powered@NLHPC: This research was partially supported by the supercomputing infraestructure of NLHPC. The work was supported by the North-German Supercomputing Alliance (HLRN). DRGS and CAO thank for funding via Fondecyt Regular (project code 1201280). PJK acknowledges the financial support by the Deutsche Forschungsgemeinschaft Heisenberg programme (grant No.\ KA 4825/4-1). FHN acknowledges funding from the DAAD.

\end{acknowledgement}


\bibliographystyle{baaa}
\small
\bibliography{687_v2}
 
\end{document}